\documentclass[aps,prd,amsmath,twocolumn,nofootinbib,longbibliography]{revtex4-1}

\usepackage{graphicx,amsmath,latexsym,color} 
\usepackage{subcaption}

\def\be{\begin{equation}}
\def\ee{\end{equation}}

\newcommand{\eq}[1]{Eq.~(\ref{#1})}

\newcommand{\boldtau}{\mbox{\boldmath$\tau$}}
\def\bea{\begin{eqnarray}}
\def\eea{\end{eqnarray}}\def\a{\alpha}
\def\g{\gamma}\def\t{\tau}\def\l{\lambda}\def\o{\omega}

\def\la{\langle}\def\ra{\rangle}\def\d{\delta}\def\bfr{{\bf r}}\def\k{\kappa}\def\b{\beta}
\def\o{\omega}
\def\r{\rho}

\def\D{\Delta}
\def\L{\Lambda}
\def\w{\omega}\def\O{\Omega}
\def\SRC{\rm{ SRC}}\def\bfr{{\bf r}}\def\bfs{{\bf s}}
 
 \def\l{\lambda}
\def\OC{\widehat{\cal O}_C(v)}
\begin{document}

\title{Nucleon-Nucleon Short-Ranged Correlations,   $\beta$ Decay and the Unitarity of the CKM Matrix  }

\author{Levi Condren and Gerald A. Miller}
 
\affiliation{
University of  Washington, Seattle  \\
Seattle, WA 98195-1560}

\date{\today}

\begin{abstract}
The influence of nucleon-nucleon short-ranged correlations (SRC) on nuclear super-allowed $\b$ decay is examined. 
The need for this is driven by  the observed depletion of spectroscopic strength obtained in  studies of $(e,e')$ and $(d,^3{\rm He})$ reactions on a wide variety of nuclei. We show that the influence  of SRC is model-dependent, but  may be very substantial. 
  The $^{46}$V nucleus is used as an example.  The resulting impact on studies of the unitarity of the  Cabibbo-Kobayashi-Maskawa  (CKM) matrix element is discussed.
\end{abstract}
 
 \maketitle

The dominant contribution to  the unitarity test of the Standard Model (SM) CKM matrix   comes  from the up-down quark matrix element $V_{ud}$.  The value of $V_{ud}$ has been  extracted by Hardy and Towner (HT)
~\cite{TOWNER197333,HARDY1975221,HARDY1990429,PhysRevC.71.055501,PhysRevLett.94.092502,1,PhysRevC.79.055502,PhysRevC.91.025501,PhysRevC.92.055505,Hardy:2020qwl} with the highest precision from $0^+\to 0^+$ decays from nuclei ranging from $^{10}$C to $^{74}$Rb. The remarkably  consistent nature of the values of $V_{ud}$ obtained from many different decays has led to a very small uncertainty. Their latest paper~\cite{Hardy:2020qwl} states
\be V_{ud}=0.97373\pm 0.00031.\label{val}\ee

Despite the  considerable success of the HT approach, the  crucial importance of the process in testing the Standard Model  has long mandated that the theory behind the analysis be continually re-examined, an especially urgent process now
  because a more recent evaluation~\cite{Shiells:2020fqp} of an electro-weak radiative correction claims a 4 standard deviation violation of  unitarity. Our focus is on the isospin-breaking correction $\d_C$.  A variation of this quantity, $\D\d_C$  would cause a change in $V_{ud}$ given by
  \be {\D (V_{ud}^2)\over V_{ud}^2}\approx \D\d_C.\label{dv}\ee
 Consider the result $\d_C=0.960(63)$\% for the $0f_{7/2}$ orbital of $^{42}$Ti~\cite{Hardy:2020qwl}. A 20 \% change, for example, in that number is about 0.2\%  and $V_{ud}$ would be changed by half that,  $ 10^{-3}$, a number that is 3.5 times the  uncertainty quoted in \eq{val}.  The Particle Data Group~\cite{Zyla:2020zbs} finds a similar central value of $V_{ud}$ but a smaller uncertainty of $\pm0.00014$. In that case the 20\% change in $\d_C$ would be almost 8 times the uncertainty.
 
 The purpose of this paper is to argue that the influence of short-ranged correlations between nucleons, unaccounted for by  Towner \& Hardy~\cite{1} (TH), may  cause  changes in the value $\d_C$, that are large on the scale of the desired accuracy.  This means that, depending on future theoretical and experimental work, either  the uncertainty in the value of  $V_{ud}$ is significantly larger than that of  \eq{val}, or that  the value is itself changed  significantly. 

Superallowed $\b$ decays are generated by the isospin operator $\boldtau$  obeying the usual commutation relations.  The theoretical formalism of TH  is based on using a weak interaction operator  different than $\boldtau$,  that does not obey these commutation rules~\cite{3,4}. The operator of TH was  designed to reduce the size of the necessary  small shell model space. 
Corrections to the TH formalism based on the collective  isovector monopole state were presented in~\cite{Auerbach:2008ut,Auerbach:2021jyt}. Work on the effects of short-ranged correlations appears in~\cite{PhysRevC.87.054304}  that concludes, ``we present a new set of isospin-mixing corrections $\cdots$, different from the values of Towner and Hardy. A more advanced study of these corrections should be performed."

The TH restriction   is motivated by a shell-model picture in which radial excitations of energy  $2\hbar \w$ and higher above the relevant orbitals can be  neglected. This approach specifically eliminates the influence of short-ranged nucleon-nucleon correlations that involve nucleons in orbitals high above the given shell model space. This strong interaction effect reduces the probability that a decaying nucleon is in a valence single-particle  orbital and suggests that the magnitude of $\d_C$ is smaller than that of previous calculations.

An exact formalism for evaluating $\d_C$ was presented in ~\cite{3,4}. 
The present effort presents an extension of that formalism  focusing on the influence of short-ranged correlations,   now known to be important because of recent significant experimental and theoretical work.


 In the time since TH started their epic sequence of calculations many new experimental and theoretical results have
obtained unambiguous evidence that nucleon-nucleon short-ranged correlations do exist in an observable fashion  
~\cite{Subedi:2008zz,Fomin:2011ng,Hen:2014nza,Hen:2016kwk,Weiss:2016obx,Stevens:2017orj,Fomin:2017ydn,CiofidegliAtti:2017xtx,Wang:2017odj,Weiss:2018tbu,CLAS:2018xvc,CLAS:2018yvt,Paschalis:2018zkx,Lynn:2019vwp,Ryckebusch:2019oya,Xu:2019wso,Lyu:2019bxr,CLAS:2020mom,CLAS:2020rue,Weiss:2020bkp,Segarra:2020plg,Shang:2020tvy,Aumann:2020tcq,Tropiano:2021qgf,Guo:2021zcs,Lu:2021xvj,Ydrefors:2021mky,Wang:2021okr}.
The  effects of short-ranged correlations between nucleons, predicted   long ago, have finally  been  measured and are significant. Such correlations involve the excitations of nucleons to intermediate states of high energy.  Consequently, radial excitations are now known to be important in nuclear physics. 

Spectroscopic factors, essentially
the occupation probability of a single-particle, shell-model orbital, play an important role in what follows. As  reviewed in Ref.~\cite{Hen:2016kwk}, electron scattering experiments typically observe only about 60-70\% of the expected number of protons.  This depletion of the spectroscopic factor was observed over a wide range of the periodic table at relatively low-momentum transfer for both valence nucleon knockout using the  $(e,e'p)$  reaction ~\cite{Lapikas:1993uwd} and stripping using the  $(d,^3He)$  reaction \cite{Kramer:2000kc}.  The missing strength of 30\%-40\% implies the existence of collective effects (long-range correlations) and short-range correlations in nuclei.
See also the substantial  theoretical analyses~\cite{Atkinson:2019xtc,Barbieri:2009md,Geurts:1996zza,Radici:2003zz,Radici:2003zz} that used detailed many-body evaluations to find that including the effects of both long and short-range correlations must be included to reproduce the results of experiments that measure spectroscopic factors. 

Ref.~\cite{Paschalis:2018zkx} made a quantitative effort to analyze the separate long (LRC) and short range (SRC) contributions to the quenching of the spectroscopic factors. Their result is the SRC contribution amounts to $ 22\%~\pm~8\%$ and the LRC contribution to $ \delta = 14\%~\pm~10\%$. This is in accordance with expectations \cite{Hen:2016kwk,CLAS:2018yvt, Fomin:2011ng,Subedi:2008zz,  Hen:2014nza} and with the results of~\cite{Aumann:2020tcq,PhysRevC.41.R24,Atkinson:2019xtc,Barbieri:2009md,Geurts:1996zza,Radici:2003zz}.  In the following we argue that, in analogy with the $(e,e'p)$ and $(d,^3He$) reactions, the superallowed beta decay measurements are impacted by the short-ranged correlations that reduce the spectroscopic strength by about 20\%.

Therefore  we re-examine the calculations of superallowed beta decay rates with an eye toward including the effects of  short-ranged correlations absent in the  TH formalism.  Doing this precisely requires separating the long range correlations inherent in the shell model of TH from the missing short range correlations. This  challenging task leads to the 
 goal of first  providing a plausibility argument, rather than a detailed evaluation. We rely on simple arguments,   starting from the basics. 

The shell  model is the starting point for  nuclear physics. In its simplest form,   the 
  nucleons are in single particle orbitals and  the $\beta$ decay matrix element   is simply an overlap between neutron and proton wave functions. If  the Hamiltonian commutes with all components of the isospin operator, the spatial overlap would be unity.
But the non-commuting interactions, such as Coulomb interaction  and the  nucleon mass difference cause the overlap to be less than  than unity. 
This leads to a non-zero value of  the isospin correction known as $\d_C$.

There must be  a further modification of the value of the matrix element because there is no fundamental single-nucleon, mean-field potential in the nucleus.  The mean field that binds the orbitals is only a first  approximation to nuclear binding. The mean-field arises from the average of two- (or more) body interactions,  but residual two- (or more) nucleon effects must remain.  There are residual interactions that cause  long-range correlations, such as particle-vibration coupling and those that cause the short-ranged correlations mentioned above.
 
 The fundamental theory for the Fermi interaction of proton beta decay involves the  isospin operator $\tau_+$
and the Fermi matrix element is then given by
$
M_F = \langle f | \tau_+ | i \rangle \,, 
$
 $|i\rangle$ and $|f\rangle$ the exact initial and final 
eigenstates of the full Hamiltonian $H=H_0+V_C$, with energy
$E_i$ and $E_f$, respectively and $V_C$ denotes the sum of
{ all} interactions that do not commute with the vector 
isospin operator.

Here we extend the  formalism of Refs.~\cite{3,4} by first developing an effective $\b$-decay one-body operator that includes the
dominant isospin-violating effects and then evaluating its matrix element in a strongly-correlated system. 
Consider single-particle proton $p$ and neutron $n$ orbitals denoted by $|v,p\ra$ and $|v,n\rangle$, in which the index $v$ denotes the space-spin quantum numbers. These are eigenstate of a Hamiltonian, $h=h_0 +U_C(p)$ with a Coulomb potential, $U_C(p) $ that acts only on  protons.   The eigenkets of $h_0$ are denoted with rounded brackets and those of $h$ with the usual Dirac notation.
Then using Wigner-Brillouin perturbation theory in $U_C$ one has:
\bea |v,p\ra=\sqrt{Z_C}|v,p) +{1\over E_v-\L_v h_0\L_v}\L_v U_C|v,p\ra
\label{pw}\eea
with $Z_C= 1-\la v,p| U_C {1\over (E_v-\L_v h_0\L_v)^2} U_C  |v,p\ra$ and $(v,(n,p)|\L_v=0$ with $|v,n\ra=|v,n)$.

The single-particle super-allowed beta decay matrix element, $M_{\rm sp}\equiv  (v,n|\t_+|v,p\ra$ is given by the overlap $(v|v\ra$:
\bea M_{\rm sp}=\sqrt{Z_C} , 
\eea
with  the proton to neutron matrix element of $\t_+$  evaluated as unity. Evaluating $\sqrt{Z_C}$ to second-order in $U_C$ leads to
\bea M_{\rm sp}\approx 1-{1\over2}(v|U_C {1\over (E_v-\L_v h_0\L_v)^2}\L_v U_C|v), \label{ob}
\eea
with 
 the second term as  the isospin correction.  This result repeats the well-known results that the electromagnetic corrections are of second-order~\cite{3,4,PhysRevLett.4.186,PhysRevLett.13.264}.
  The dominant isospin correction of TH    is  twice the second term.

Next we turn to nuclear super-allowed $\b-$decay. It is useful to define  the one-body Coulomb-correction operator that appears in \eq{ob} as  $\widehat{\cal O}_C(v)\equiv U_C {1\over (E_v-\L_v h_0\L_v)^2}\L_v U_C$.
Consider, as  a first step,  a simplified situation
in which the initial nucleus $i$ consisting of a   proton in a valence orbital $v$ outside an isospin-0 core state of $A$ nucleons  beta decays to a neutron outside the same  state, $f$. Then the  $p\to n$ matrix of $\t_+$ is still unity and is not mentioned below.
The core of the state $f$ is taken to be the same as that of the state $i$, so that their overlap  does  not influence the $\b$ decay matrix element. This is an accurate treatment  because the only isospin-violating influence on the wave function is  caused by the external valence proton, a negligible ${\cal O} (1/A)$ effect. Then the largest Coulomb correction is obtained by
taking the matrix element  $\delta_C(v)$ of the operator $\widehat{\cal O}_C(v)$. In coordinate-space and suppressing spin indices, this quantity is given by:
\be 
{\d_C}_0(v)=\int d^3rd^3r' \phi^*_v(\bfr) {\cal O}_C(\bfr,\bfr') \phi_v(\bfr').\label{m01}\ee

The simple single-particle state leading to \eq{m01} is only a first, mean-field  approximation to the nuclear wave function. This is because
the valence proton (neutron) undergoes strong interactions with the core nucleons that  involve both
long- and  short-ranged correlations.  Focussing on short-ranged, two-nucleon aspects,
we need to compute the two-nucleon wave function given by
\bea |v,\a\ra=\sqrt{Z_S(v,\a)}|v,\a) +Q  {G\over e}|v,\a),\label{form1}\eea
 with $\a$ being one of the occupied orbitals of the $T=0$ core state. Here $G$ \footnote{For two-nucleon interactions $G$ is the two-nucleon $T$-matrix evaluated at negative energy and modified by Pauli blocking effects.} is the 
 anti-symmetrized~\cite{Hugenholtz:1957zz} reaction matrix operator  that sums ladder diagrams involving 
two--nucleon interactions. The factor  $Z_S$   insures the normalization,  $e$  represents  an energy denominator and  iterations of the potential that correct the state $|i_0)$ are included in the schematic factor $Q {G\over e}$
The Hermitian projection operator $Q$ obeys $Q|v,\a)=0$.
 and is constructed  to exclude the long-ranged correlations  so that {\it only} the short-ranged correlations are included in the correction we study.  \eq{form1} includes only the leading-order term in the linked-cluster expansion of 
 Refs.~\cite{Goldstone:1957zz,Hugenholtz:1957zz,Brandow:1967fbn,Shakin:1971zz,daProvidenica:1971zmp}.
 Defining  an operator $\Omega\equiv Q {G\over e}$, one has  $Z_S(v,\a)=1-(v,\a|\O^\dagger\O|v,\a)$. Then
\begin{widetext}
 \bea \d_C(v)=Z_S(v)(v|\OC|v) +
 \sum_\a [(v,\a|\sqrt{Z_S(v,\a)}  (\OC\O+\O^\dagger \OC) +\O^
 \dagger\OC\O|v,\a)]
. \label{dcv}\eea
\end{widetext}
with
$Z_S(v)\equiv {}1-\sum_\a (v,\a|\O^\dagger\O|v,\a)	\equiv1- \k(v)
.$ This is the occupation probability, known always to be $<1$. 

In this first analysis of the effect of $\SRC$ on super-allowed $\b$-decay we rely on the existing literature that indicates that 
$\, Z_S(v)\approx 0.8$ for many states $v$, although with dependence on the specific state, nucleus and interactions.  This number comes from many experimental measurements and theoretical calculations cited  above.  To be specific, consider
the case of a single proton  in a  $0f_{7/2}$ state outside an inert  core of charge $Z=22$, schematically  representing the calculation for $^{46}  $V. A search of the literature~\cite{Aumann:2020tcq} for $Z_S(0f_{7/2})$ reveals a value of $\k(0f_{7/2})=0.14$ for that state in $^{55}$Ni. This is for a Hamiltonian with  ``mild short-range 
repulsion effects". This value is consistent with the results of the independent phenomenological analysis of~\cite{Paschalis:2018zkx}. In the following we use 
$\k=1-Z_S$ to simplify the notation.

Next turn to an evaluation of the matrix element $(v|\OC|v) $ that provides a proof of the validity of \eq{m01}.
 Our numerical results and  Hartree-Fock 
 calculations~\cite{PhysRevC.1.1260} show that the valence radial wave function is very well approximated by that of a three-dimensional harmonic oscillator. We therefore
use
harmonic-oscillator single-particle wave functions with the parameters of ~\cite{PhysRevC.1.1260}.

The nuclear Coulomb potential arises from the convolution of 
$Z \a /( |\bfr-\bfr'|)$ with the charge density $\rho_C(r')$. If we take the latter to be a constant within $r \le R_C$,
the one-body Coulomb potential
takes the form used by TH.
The value of $R_C$ is chosen to match the Coulomb potential obtained with a Fermi shape using $R_A=1.1 A^{1/3}$fm and  $a=0.54$ fm.
Our  estimate takes the state $|v)$ to be in the single-particle orbit with
radial quantum number $n=0$ and angular momentum $l=3$ appropriate for the state appearing in the first line of Table I for $^{46}V$ of Ref.~\cite{1}. 
The matrix element of $U_C$ between the valence state and the state with $n,l$ is $(0l\bigl|U_c\bigr|nl)$
so that using
\eq{m01}, we find
\be
{\delta_C}_0(l) =  \sum\limits_{n > 0}
\, \frac{\biggl|(0l\bigl|U_c\bigr|nl)\biggr|^2}{4n^2\w^2}.
\label{dcalc}\ee
Using this yields  ${\d_C}_0=0.267\,$\% in agreement with the result  
in  Table I for $^{46}V$ in Ref.~\cite{1}.

At this stage the result is that the leading Coulomb correction of TH is multiplied by the factor $Z_S$, potentially
 a very substantial reduction in terms of present accuracy requirements

Now we turn towards  the remaining terms of \eq{dcv}.
First note  that the operator $ \O$ contains the projection operator $Q$ that projects away from the initial state. The mean-field state $|v\a)$ is  changed
by the action of $QG$ to one in which one  or  mainly both nucleons are above the Fermi sea. Then the one-body operator $\OC$ cannot connect the intermediate state to the mean-field state.
Thus the terms   of \eq{dcv} that are linear in the operator $\O$ vanish and one has:
\bea &\d_C(v)=Z_S(v)(v|\OC|v) +
 \sum_\a (v,\a|\O^
 \dagger\OC\O|v,\a)
. \nonumber\\&\label{dcv1}\eea

Next we estimate the matrix elements of $\O^
 \dagger\OC\O$, the second-order terms. 
This is implemented here  by modeling  the operator $\Omega$
using  the Jastrow correlation~\cite{PhysRev.98.1479,Feshbach:1958nx,Czyz:1969jg,Miller:1975hu,Pandharipande:1979bv,Mahaux:1985zz,Brockmann:1990cn,Engel:2011ss,Shakin:1971zz,daProvidenica:1971zmp} approximation to the   nuclear wave function.
The most important and best-measured $\SRC$ involve two  nucleons, in which the correlated wave function, $\psi^{(2)}$ is related to the  mean field approximation $\phi^{(2)}$ with $\psi^{(2)} =(1+f)\phi^{(2)} $.  In the present situation one of the nucleons is the decaying proton in orbital $v$   and the other is any nucleon in  the occupied orbital $\a$, so $\phi^{(2)}$ represents the product state $|v\a)$. The correlations (including Pauli) are
 represented by a function $f(s)$, in which $s$ is the separation distance.   
 A schematic notation, in which various quantum numbers of the two-nucleon wave function are not explicitly specified, simplifies the  presentation.  Then, the operator $\Omega$ and $f(s)$ are related by $f(s)=\la s|\Omega|\phi^{(2)}\ra$~\cite{Shakin:1971zz,daProvidenica:1971zmp}. The function $f(s)$ is meant to  represent {\it only} the short-ranged correlations, as mandated by the proper construction of the operator $Q$. The operator $\O$ acts only in two-nucleon states allowed by the Pauli principle. 
A first-principles calculation would explicitly state angular momentum-dependence $LJST$  dependence of the functions $f$. Indeed several partial-wave contributions enter in the computation of $\k=\sum_v\k(v)/A$~\cite{Preston}.

The nuclear force is repulsive at small separations and attractive at large separations. This means that, in all existing models, $f$ is negative for small values of $s$, rises to 0 or slightly above  as  values of $s$ increase towards  the region of attraction and then falls to within 1 fm or so~\cite{Miller:1975hu,Engel:2011ss,Chen:2016bde,PhysRevC.72.054310,PhysRevC.79.055501,PhysRevC.72.034002,Cruz-Torres:2017sjy}.
 The details of $f(s)$ are model dependent, but the previous sentence  holds. The function $f(s)$ is substantial only for small values of $s$. 

Next we use the short-ranged nature of $f(s)$ to compute
  the second-order terms of \eq{dcv1}. Defining that term  as $ \D{\d_C}_0(v)$, and evaluating in coordinate-space leads to the expression:
 \be \D{\d_C}_0(v)=\int d^3rd^3r'
\phi_v(\bfr)I(\bfr,\bfr')
{\cal O}_C(\bfr,\bfr')\phi_v(\bfr')
,\label{big}\ee
with 
\bea&
&I(\bfr,\bfr')\equiv\int d^3r_2\r(\bfr_2) f(|\bfr-\bfr_2|)f(|\bfr'-\bfr_2|).\label{I}\eea
  
An examination of the integrals shows that $ \D{\d_C}_0(v)>0,$ and tends to compensate for the reduction caused by using $Z_(v)<1.$
A simple general analysis that focuses on the short-distance repulsion is used.
Let us first suppose that the short ranged correlations are captured by using $f(s)=-\l g(s)$, with $0< \l\le 1$  and $g(0)=1$, and $g(s) $ is vanishing for 
values of $s$ larger than a reasonable range,  $r_0$, of order 1 fm or less.
  This form is a reasonable and flexible representation of the short distance properties  all of the models~\cite{Miller:1975hu,Akmal:1997ft,Cruz-Torres:2017sjy,PhysRevC.72.054310,PhysRevC.79.055501,PhysRevC.72.034002,PhysRevC.51.38,PhysRevC.63.024001} that allows us to study the model dependence.
  The values of $\l$ and the function $g(s)$ are chosen to reproduce the value of $\k$ via 
  \bea 
  \k =\r_0 \l^2 \int d^3 s\, g^2(s) 
  \label{kr},\eea
  where $\r_0$ is the density of nuclear matter $\approx 0.167 \,\rm fm^3$.  
  Our philosophy is to take the value of $\k$ as determined by experimentally measured spectroscopic factors and independent theory.
  
  For the $^{46}$V example used here, $\k=0.14$. Then using a Gaussian form, $g(s)=e^{-s^2/2r_0^2},$ leads to $r_0=0.532/\l^2$ fm, and using a square shape,
  $ g(s)= \Theta(r_0-s),$ leads to
  $r_0=0.585/\l^2$ fm.
  
  Now turn to the evaluation of $I(\bfr,\bfr') $. 
  Note that because of the short-ranged nature of the correlations
  the integrand of $I(\bfr,\bfr') $ is substantial  only when both $\bfr$ and $\bfr'$ are close to $\bfr_2,$ and therefore close to each other.  
  Because $r_0$ is much less than the nuclear radius we approximate as a three-dimensional delta function via:
  \bea g(s) =\d(\bfs)\,\int d^3s g(s)\,.\label{df}\eea
   Numerical analysis shows that using this simplification provides an excellent approximation to the exact calculation.
 The ratio  \bea  
 \g\equiv {\int d^3s \,g(s)\over\int d^3s \,g^2(s)}
  \eea
  is an important parameter in the following treatment.

Using \eq{df} and \eq{kr} to evaluate $I(\bfr,\bfr')$ leads to the result
\bea
I(\bfr,\bfr')\approx{\g^2\over \l^2}\Big( { \k\over\r_0}\Big)^2\r(\bfr)\d(\bfr-\bfr')
.\label{IA}\eea  
Using this expression to compute $\D{\d_C}_0(v)$ of \eq{big} leads to:
\bea
&\d_C(v)=Z_S(v) {\d_C}_0(v)\nonumber\\&+{\g^2\k^2\over \l^2\r_0^2}\int d^3r \phi_v^*(\bfr){\cal O}_C(\bfr,\bfr')\phi_v(\bfr')\r(\bfr)
\label{tot}\eea
 showing that the second-order term tends to compensate for the depletion caused by the factor $Z_(v)<1$. Furthermore, there is 
strong sensitivity to the value of $\g\k/\l$.  

Next use   \eq{IA},  to compute $\D{\d_C}_0(v)$ of \eq{big} that is also the second term of \eq{tot}. The $^{46}$V model and parameters used to compute $\d_{{C}_0}$.  is again used. Then
\bea&
\D{\d_C}_0(v)={\g^2\over \l}\Big( { \k\over\r_0}\Big)^2{2l+1\over 4\pi}\sum_{n=1}{\int r^2 dr R^2_{0l}(r)R^2_{nl}(r)U_C^2(r)
\r(r)\over 4n^2\o^2}\nonumber\\&={\g^2\over\l^2}\Big( { \k\over\r_0}\Big)^2{2l+1\over 4\pi} 0.087 {\d_C}_0,\eea
numerical evaluation and the results of using \eq{dcalc}. For the case of interest ($l=3, \k=0.14$) we find
\bea
\D{\d_C}_0(v)={\g^2\over \l^2} 0.034 \,{\d_C}_0.
\eea
The value of $\g$ is determined by the shape of $g(s)$. Using a square shape yields $\g^2=1$ and using a Gaussian yields $\g^2=8.$ With the former (and $\l=1$) one finds that $\D{\d_C}_0(v)$ is negligible, but with the latter the correction is $0.27 {\d_C}_0$ and the effects of short-ranged correlations is to provide an overall increase of about 13\%. One may use a Fermi function $g(s)= 1/(1+\exp(s-r_0)/a)$. In that case varying the value of $a$ from small values to about 1.9 fm smoothly interpolates the values of  $g^2$  between 1 and 8. For the specific example, one obtains either an 11 \% decrease with $\g^2=1$ or a 13\% increase with $\g^2=8$. If the value of $\k$ is taken to be 0.2 for stronger short-range repulsion this spread goes from a 20\% decrease to a 30 \% increase.  More generally:   the resulting electromagnetic corrections to super allowed beta decay can be increased or decreased substantially by the influence of short-ranged correlations. 




 Our considerations are limited to one state. The nuclear dependence of $\d_C$ is important as established by TH. Roughly speaking, our trend is very similar to theirs because the driving effect is the increase of the Coulomb interaction with increasing nuclear size. The cited theory and measurement references on the $A$-dependence of spectroscopic factors indicate that the influence of  SRC is likely to have little A-dependence. Thus trends similar to that of TH are to be expected.

 We summarize.
The key result is that computations of the isospin correction are strongly sensitive to the effects of short-ranged correlations.  The detailed features of the short-ranged correlations determine whether the influence is an increase, decrease or no change. This is true despite the schematic nature of the present calculations. The correct  evaluation of this effect can only be assessed precisely by doing detailed calculations with different models that account for the experimentally measured spectroscopic factors. This is important because tests of the unitarity of the CKM matrix demands very high accuracy.
Doing more detailed state-of-art nuclear calculations of superallowed $\b$ decay is a high priority for nuclear theorists.

{\bf Acknowledgments}
This work was supported by the U.S. Department of Energy Office of Science, Office of Nuclear Physics under Award No. DE-FG02-97ER-41014.
We thank A.~Garcia, O.~Hen, A. Schwenk, S.~R.~Stroberg and U.~van~Kolck for  useful discussions.
%

\end{document}